\documentclass[runningheads]{llncs}

\usepackage[T1]{fontenc}
\usepackage{graphicx}
\usepackage{xcolor}
\usepackage{comment}
\usepackage{amssymb}
\usepackage{amsmath,bm}
\usepackage{tabularx}
\usepackage{makecell}
\usepackage{rotating}
\usepackage{threeparttable}
\usepackage{booktabs}
\usepackage{array}
\usepackage{multirow}
\usepackage{subfigure}
\usepackage{verbatim}
\usepackage{lineno}
\usepackage{hyperref}

\begin{document}

\title{SKAD: Sketch-Guided Audio-Contrastive Decoding for Reasoning in Large Audio-Language Models}

\titlerunning{SKAD: Sketch-Guided Audio-Contrastive Decoding for
Reasoning in LALMs}

\author{Junyu Wang\inst{1} \and
Peiyuan Jiang\inst{1} \and
Siyuan Zhang\inst{1} \and
Jian Zong\inst{1} \and
Tianrui Wang\inst{1} \and
Zikang Huang\inst{1} \and
Meng Ge\inst{1} \and
Longbiao Wang\inst{1,2}\thanks{Corresponding author.} \and
Jianwu Dang\inst{3}}

\authorrunning{J. Wang et al.}

\institute{Tianjin Key Laboratory of Cognitive Computing and Application, College of Intelligence and Computing, Tianjin University, Tianjin, China\and
Huiyan Technology Co., Ltd, Tianjin, China\and
Shenzhen Institute of Advanced Technology, Chinese Academy of Sciences, Guangdong, China\\
\email{junyu\_wang21@tju.edu.cn, longbiao\_wang@tju.edu.cn}
}


\maketitle

\begin{abstract}
Large Audio-Language Models (LALMs) have shown increasing potential in audio understanding and reasoning by generating intermediate reasoning steps before answering. However, directly eliciting long-form reasoning from audio does not always lead to more reliable predictions. Even reasoning-oriented LALMs may gradually shift from acoustic evidence to language-continuation patterns during generation. A natural solution is to use contrastive decoding with an audio-masked branch, but directly applying this strategy to plain chain-of-thought reasoning is often ineffective, since the masked branch lacks sufficient audio-related context and provides an unstable reference. 
To address this issue, we propose \textbf{SKAD} (\textbf{Sk}etch-Guided \textbf{A}udio-Contrastive \textbf{D}ecoding), a training-free inference-time method for audio reasoning. SKAD first constructs an acoustic evidence sketch that summarizes question-relevant auditory information. It then performs contrastive decoding between an audio-conditioned reasoning branch and a sketch-only branch. By using the sketch as a shared semantic scaffold, SKAD makes the masked branch more informative and encourages decoding decisions that are more dependent on the original audio. 
We evaluate SKAD on Qwen2.5-Omni-7B and Step-Audio-R1 across three audio understanding and reasoning benchmarks: MMAR, MMAU, and MMSU. Experimental results show that SKAD consistently improves the average accuracy of both models across all benchmarks. Further ablations show that direct contrastive decoding without the acoustic evidence sketch provides limited or negative gains, confirming the importance of constructing a reliable contrastive reference before audio-contrastive decoding.
Code is available at \url{https://github.com/MyParadise21/SKAD}

\end{abstract}

\section{Introduction}
\label{intro}

Large Audio-Language Models (LALMs) have become a general interface for understanding speech, environmental sounds, music, and mixed acoustic scenes \cite{qwen2audio,Audio-Flamingo2,Audioflamingo3}. Recent models are increasingly expected to answer questions that require not only acoustic perception, but also reasoning over the perceived evidence \cite{qwen25omni,kimiaudio,mimoaudio}. For example, a model may need to compare sound events, infer speaker or scene attributes, identify temporal relations, or combine multiple acoustic cues before selecting an answer. This makes audio understanding different from simple event recognition: the model must generate responses that are both semantically plausible and supported by the input audio.

Despite recent progress, reliable audio reasoning remains challenging. Audio signals are continuous, transient, and often only partially aligned with textual concepts. During multi-step generation, an LALM may produce reasoning that sounds coherent but is weakly tied to the acoustic evidence \cite{stepaudior1,qwen3omni}. In particular, the model can continue from its own textual context, rely on common-sense priors, or introduce plausible intermediate claims that are not directly supported by the audio. This reflects a general difficulty when language models are used to reason over non-textual evidence. Contrastive decoding provides a natural inference-time mechanism for improving modality grounding \cite{contrastivedecodellm}. In multimodal generation, a common strategy is to compare a modality-conditioned branch with a modality-masked branch, and then promote tokens that are more strongly supported by the input modality \cite{contrastivemllm}. When applied to audio reasoning, one may directly contrast an audio-conditioned chain-of-thought branch with an audio-masked chain-of-thought branch. However, we find that this direct adaptation is ineffective. Once the audio is removed at the beginning of reasoning, the masked branch has little access to task-relevant acoustic information, while still being required to generate a complete reasoning process. Its token distribution therefore tends to reflect arbitrary guesses, generic reasoning patterns, or unstable language priors, rather than a meaningful reference for contrastive decoding. We refer to this problem as masked-branch degeneration.

This observation suggests that audio-contrastive decoding requires a shared evidence context before the audio-masked branch can serve as a useful reference. To this end, we propose \textbf{SKAD}, a training-free sketch-guided audio-contrastive decoding method. Given an audio input and a question, SKAD first prompts the model to generate a compact acoustic evidence sketch that summarizes task-relevant acoustic cues. The sketch is then provided to two parallel reasoning branches. The audio-conditioned branch receives both the original audio and the sketch, while the sketch-only branch receives the same sketch with the audio information masked. Since both branches share the same intermediate evidence, the sketch-only branch is no longer forced to reason from an empty acoustic context; instead, it estimates how the model would reason from the textualized audio evidence alone. During the subsequent decoding process, SKAD compares the token distributions of the two branches. At most reasoning steps, the two distributions are largely aligned and no substantial correction is needed. The contrastive signal becomes useful at token positions where the two branches diverge: in such cases, the sketch-only branch may rely mainly on the textual sketch or general language priors, while the audio-conditioned branch can reflect additional acoustic cues that are not fully captured by the sketch. We therefore reshape the decoding logits at these informative positions to strengthen audio-discriminative predictions from the audio-conditioned branch.

SKAD does not require additional training data, external models, parameter updates, or architectural modifications. It only changes the inference-time decoding process and can be applied to LALMs that expose token-level prediction distributions. Compared with original chain-of-thought reasoning, the proposed framework introduces an explicit evidence sketch before reasoning. Compared with directly applying contrastive decoding to original chain-of-thought generation, it avoids constructing the negative branch from a severely underconstrained audio-masked prompt. We evaluate SKAD on widely used audio understanding and reasoning benchmarks, including MMAU, MMAR, and MMSU \cite{mmau,mmar,mmsu}. Experimental results show that SKAD improves average accuracy over direct chain-of-thought baselines across the evaluated benchmarks. Ablation studies further show that contrastive decoding is more effective after a shared evidence sketch is established, suggesting that successful audio contrastive decoding depends not only on masking the audio modality, but also on constructing a meaningful reference branch for comparison.

Our main contributions are summarized as follows:
\begin{itemize}
    \item We identify masked-branch degeneration as a practical limitation of directly applying contrastive decoding to audio chain-of-thought reasoning, where the audio-masked branch lacks sufficient task-relevant evidence to provide a useful reference.
    \item We propose SKAD, a training-free sketch-guided audio-contrastive decoding method, which uses an acoustic evidence sketch as a shared context for comparing an audio-conditioned branch with a sketch-only branch.
    \item We conduct experiments and ablations on MMAU, MMAR, and MMSU, showing that SKAD improves over direct chain-of-thought baselines and that contrastive decoding is more effective after a shared acoustic evidence sketch is established.
\end{itemize}

\section{Related Work}

\subsection{Large Audio-Language Models and Audio Reasoning}

Large Audio-Language Models (LALMs) integrate acoustic encoders with instruction-following language models, enabling language-based interaction with speech, environmental sounds, music, and mixed audio inputs. Recent systems have shown strong performance on audio captioning, audio question answering, speech understanding, music analysis, and general audio instruction following \cite{LTU,salmonn,baichuanomni,stepaudio2}. As these models become more capable, evaluation has also moved from recognition-oriented tasks toward questions that require models to select, combine, and reason over auditory cues. Benchmarks such as MMAU, MMAR, and MMSU provide such evaluation settings across speech, sound, music, mixed audio, and spoken language understanding \cite{mmau,mmar,mmsu}. In parallel, recent studies improve audio reasoning by introducing intermediate reasoning traces, constructing reasoning-oriented data, applying reinforcement learning, or strengthening modality-aware supervision \cite{Audio-cot,Audio-reasoner,Audio-Thinker,stepaudior1}. These methods mainly enhance the model through prompts, data, or training objectives. Our work takes a different route: we keep the underlying LALM fixed and study how inference-time decoding can make the reasoning process more sensitive to the audio input.

\subsection{Chain-of-Thought Reasoning and Training-Free Decoding}

Chain-of-thought (CoT) prompting encourages language models to produce intermediate steps before final answers and has been widely used for complex reasoning \cite{COT1,deepseekR1}. For audio tasks, intermediate reasoning can provide a useful textual workspace for organizing perceived events and attributes. However, the same generation process may also be influenced by textual continuation patterns, especially when the model must elaborate beyond clearly perceived acoustic cues. Training-free decoding methods offer a complementary way to adjust generation without updating model parameters. Contrastive decoding compares token distributions from different models, layers, or input conditions to improve generation quality or reduce undesired priors \cite{contrastivedecodellm}. In multimodal generation, related methods contrast modality-conditioned predictions with predictions under weakened or perturbed inputs to mitigate hallucination and improve grounding \cite{contrastivemllm,icd}. Recent audio-language studies also explore inference-time interventions that increase the influence of audio information during generation \cite{mata}. Different from these settings, we find that directly using the modality-masked branch as a contrastive reference in audio chain-of-thought reasoning suffers from masked-branch degeneration: once the audio is removed during reasoning generation, this branch may lack sufficient task-specific content and fail to form a stable reference distribution. To address this issue, SKAD first constructs an acoustic evidence sketch and then compares the audio-conditioned branch with the sketch-only branch under a shared reasoning context. This design makes the contrastive reference more informative and better aligned with the audio-conditioned decoding process.

\begin{figure}[t]
\centering
\includegraphics[width=12cm]{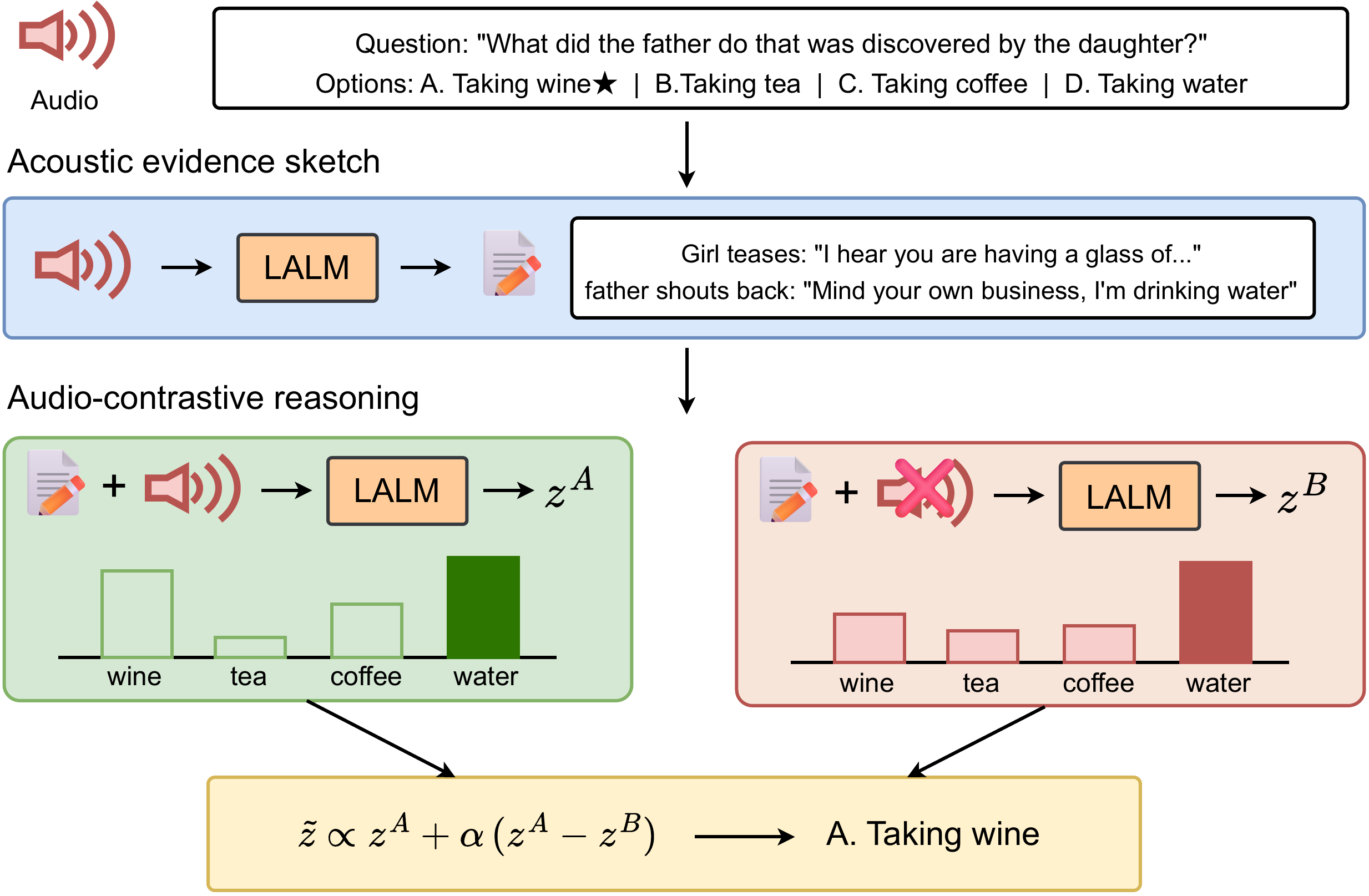}
\caption{Overview of SKAD, the proposed sketch-guided audio-contrastive decoding method. Given an audio input and a question, the model first generates an acoustic evidence sketch that summarizes task-relevant auditory evidence. The sketch is then used to condition two parallel reasoning branches: an audio-conditioned branch that receives both the sketch and the original audio, and a sketch-only branch where the audio is masked. Their token logits, denoted by $z^{A}$ and $z^{B}$, are contrasted to reshape the decoding distribution, so that predictions more strongly supported by the original audio are emphasized.}
\label{diagram1}
\end{figure}

\section{Methodology}
\subsection{Overall Framework}

Fig.~\ref{diagram1} illustrates the overall pipeline of SKAD. Given an audio input $x$, a question $q$, and optional answer candidates $c$, the goal is to generate an answer $a$ that is both semantically valid and grounded in the audio. A natural approach is to employ contrastive decoding, which compares predictions made with and without access to the original audio to emphasize information additionally supported by the audio. A straightforward way to apply contrastive decoding to audio reasoning is to run two reasoning branches from the original prompt: one conditioned on the audio and the other with the audio masked. However, without intermediate acoustic evidence, the audio-masked branch contains limited task-relevant information and may fail to provide a stable and informative reference distribution, making the resulting contrastive signal unreliable.

To address this issue, SKAD first generates a compact acoustic evidence sketch from the audio and the question, providing a partial textual representation of the auditory evidence. We then perform reasoning with two parallel branches: an audio-conditioned branch that receives both the original audio and the sketch, and a sketch-only branch that receives the same sketch while the audio is masked. Since the two branches share the same intermediate evidence, their token distributions become more comparable. The final decoding distribution is obtained by contrasting the two branches, so that predictions additionally supported by the original audio can be emphasized. The overall SKAD process can be summarized as
\begin{equation}
    audio,\, question \;\longrightarrow\; sketch \;\longrightarrow\; reason \;\longrightarrow\; answer
    \label{eq:pipeline}
\end{equation}
where $sketch$ denotes the acoustic evidence sketch.

\subsection{Acoustic Evidence Sketch}

The acoustic evidence sketch is designed to guide the model to verbalize potentially relevant auditory cues before reasoning. These cues may include sound events, speaker behaviors, dialogue fragments, scene attributes, or temporal relations, depending on the input audio and the question. Since the sketch is a compact textual representation, fine-grained acoustic evidence may still be omitted, weakened, or expressed ambiguously in text. We therefore use it as a lightweight scaffold for constructing an effective reference branch in the subsequent contrastive reasoning stage.

Formally, let $M_{\theta}$ denote the backbone audio-language model. Given the audio input $x$, question $q$, and optional answer candidates $c$, we construct a sketch-generation prompt $u_s(q,c)$ and ask the model to generate
\begin{equation}
    s \;=\; M_{\theta}\!\left( x,\; u_{s}(q, c) \right)
    \label{eq:sketch}
\end{equation}
The generated sketch is fixed during the subsequent reasoning stage. It provides the sketch-only branch with a shared textual context derived from the audio, so that this branch estimates how the model would reason from the textualized auditory evidence alone. This avoids using an unconstrained audio-masked branch as the contrastive reference while preserving a meaningful comparison with the audio-conditioned branch.

\subsection{Sketch-Guided Audio-Contrastive Decoding}

After obtaining the sketch $s$, we decode the reasoning sequence with two synchronized branches. At decoding step $t$, let $y_{<t}$ be the prefix that has already been generated. The audio-conditioned branch computes token logits with access to the original audio:
\begin{equation}
    z_{t}^{A} \;=\; f_{\theta}\!\left( \,\cdot \,\middle|\, x,\, q,\, c,\, s,\, y_{<t} \right),
    \label{eq:z_a}
\end{equation}
whereas the sketch-only branch computes token logits under the same textual context but with the audio masked:
\begin{equation}
    z_{t}^{B} \;=\; f_{\theta}\!\left( \,\cdot \,\middle|\, \varnothing,\, q,\, c,\, s,\, y_{<t} \right),
    \label{eq:z_b}
\end{equation}
where $\varnothing$ denotes the masked audio input. The two branches share the question, answer candidates, sketch, and generated prefix, and differ only in whether the original audio is available. Therefore, $z_{t}^{B}$ serves as a partial-evidence reference that captures what can be inferred from the sketch and textual context, while $z_{t}^{A}$ additionally reflects information from the original audio.

We then reshape the decoding logits by contrasting the two branches:
\begin{equation}
    \tilde{z}_{t} \;=\; z_{t}^{A} \;+\; \alpha\,\bigl( z_{t}^{A} - z_{t}^{B} \bigr),
    \label{eq:contrast}
\end{equation}
where $\alpha$ controls the strength of the contrastive adjustment. When the two branches produce similar distributions, the contrastive term remains small and the decoding stays close to the original audio-conditioned prediction. When their distributions differ, the adjustment amplifies predictions that are more strongly supported by the audio-conditioned branch, suggesting that the original audio provides additional evidence beyond the shared textual sketch.

The next token is generated from the adjusted logits using the model's original decoding rule:
\begin{equation}
    y_{t} \;=\; \operatorname{Decode}\!\left( \tilde{z}_{t} \right)
    \label{eq:decode}
\end{equation}
The generated token is appended to the shared prefix and used by both branches in the next decoding step. This synchronized decoding process continues until the model produces the final answer. Since SKAD only modifies logits at inference time, it requires no additional training data, parameter updates, external modules, or architectural changes.

\section{Experimental Setup}

\subsection{Datasets and Models}

We evaluate our method on three widely used audio understanding and reasoning benchmarks: MMAU, MMAR, and MMSU. MMAU covers diverse audio understanding tasks over speech, environmental sounds, and music, and evaluates both acoustic perception and reasoning ability. MMAR focuses on deeper reasoning over speech, sound events, music, and their mixtures, providing a challenging setting for testing whether LALMs can infer answers from multiple auditory cues. MMSU targets spoken language understanding and reasoning, involving linguistic, semantic, prosodic, and paralinguistic information in natural speech.

We conduct experiments on two representative audio-language models: Step-Audio-R1 and Qwen2.5-Omni-7B. Step-Audio-R1 is a reasoning-oriented LALM that is specifically designed to produce intermediate reasoning before giving the final answer. It therefore provides a strong testbed for evaluating whether inference-time contrastive intervention can further improve audio-grounded reasoning. Qwen2.5-Omni-7B is a general-purpose omni-modal model with audio understanding capability, but it is not specifically optimized for long-form audio reasoning. Evaluating on both models allows us to examine whether the proposed method is effective not only for a specialized reasoning model, but also for a more general omni-modal model.

\subsection{Evaluation Protocol}

For each example, the model is given the audio input, the question, and the candidate answers, and is required to generate the final answer after reasoning. We report accuracy as the primary evaluation metric. Greedy decoding is used for all methods to ensure a deterministic and comparable evaluation setting.
For SKAD, we first generate an acoustic evidence sketch and then apply audio-contrastive decoding during the reasoning stage. The contrastive strength $\alpha$ is selected according to the reasoning characteristics of each backbone model. For Step-Audio-R1, which already has strong built-in long-form reasoning ability, we use a relatively mild contrastive strength of $\alpha=0.2$ to avoid overly perturbing its reasoning process. For Qwen2.5-Omni-7B, which is a general omni-modal model rather than a reasoning-specialized LALM, we use a stronger contrastive strength of $\alpha=0.5$ to provide more explicit audio-grounding guidance during decoding. Unless otherwise specified, these settings are used across all benchmarks.

\section{Experimental Results and Analysis}
\subsection{Main Results}

\begin{table*}[t]
\renewcommand{\arraystretch}{1}
\small
\centering
\caption
{Accuracies (\%) on MMAR benchmark. S, M, and Sp denote sound, music, and speech, respectively. \textsuperscript{\textdagger} represents results reproduced in this work. \textbf{Bold} and \underline{underlined} indicate the best and the second best results among all models, respectively.}
\resizebox{\linewidth}{!}{%
\begin{tabular}{lcccccccc}
\toprule
\multirow{2.5}{*}{\textbf{Model}} & \multicolumn{3}{c}{\textbf{Single Modality }} &  \multicolumn{4}{c}{\textbf{Mixed Modalities}} & \multirow{2.5}{*}{\textbf{Avg (\%)}} \\ 
\cmidrule{2-8}
& \textbf{Sound} & \textbf{Music} & \textbf{Speech} & \textbf{S--M} & \textbf{S--Sp} & \textbf{M--Sp} & \textbf{S--M--Sp} & \\
\midrule
\multicolumn{9}{l}{\textit{\textbf{Baselines:}}} \\
Gemini 2.0 Flash & \textbf{61.2} & \textbf{51.0} & 72.1 & \textbf{81.8} & \underline{72.5} & 65.9 & 70.8 & \textbf{65.6}   \\ 
GPT-4o Audio & 53.9 & \textbf{51.0} & 70.4 & \underline{63.6} & \underline{72.5} & 62.2 & \underline{75.0} & 63.5 \\
SALMONN & 30.3 & 31.1 & 34.7 & 9.1 & 34.9 & 35.4 & 41.7 & 33.2 \\
Qwen2-Audio-7B-Instruct & 33.3 & 24.3 & 32.3 & 9.1 & 31.2 & 30.5 & 25.0 & 30.0  \\
Audio Flamingo 2 & 24.9 & 17.5 & 20.8 & 18.2 & 26.6 & 23.2 & 8.3 & 21.9  \\
Audio Flamingo 3 & \underline{58.8} & 40.8 & 59.9 & 54.6 & 61.9 & \underline{67.1} & 58.3 & 56.7 \\
\midrule
\multicolumn{9}{l}{\textit{\textbf{Ours:}}} \\
Qwen2.5-Omni-7B\textsuperscript{\textdagger} & 55.2 & 42.7 & 55.8 & 54.6 & 59.2 & 53.7 & 50.0 & 53.4 \\
\quad + SKAD & 57.0 & 43.2 & 59.9 & 54.6 & 63.3 & 58.5 & 58.3 & 56.5 \\
Step-Audio-R1\textsuperscript{\textdagger} & 52.7 & \underline{46.1} & \underline{72.8} & 45.5 & 70.2 & \textbf{68.3} & 66.7 & 62.6 \\
\quad + SKAD & 57.6 & 44.7 & \textbf{73.5} & 45.5 & \textbf{75.7} & \textbf{68.3} & \textbf{79.2} & \underline{64.8} \\
\bottomrule
\end{tabular}%
}
\label{MMAR}
\end{table*}

Tables~\ref{MMAR} and~\ref{MMAUMMSU} report the main results on MMAR, MMAU, and MMSU. Across all three benchmarks, SKAD consistently improves the average accuracy of both backbone models. For Qwen2.5-Omni-7B, SKAD brings gains of 3.1 points on MMAR, 3.4 points on MMAU, and 3.7 points on MMSU. For Step-Audio-R1, the corresponding gains are 2.2, 3.7, and 2.3 points, respectively. These results show that SKAD is effective for both a general omni-modal model and a reasoning-oriented audio model.

On MMAR, SKAD improves Qwen2.5-Omni-7B from 53.4\% to 56.5\%, and improves Step-Audio-R1 from 62.6\% to 64.8\%. The gains are more evident on several mixed-modality subsets, suggesting that SKAD is helpful when the answer depends on multiple types of auditory cues. On MMAU Test-mini, SKAD improves Qwen2.5-Omni-7B from 70.3\% to 73.7\%, and improves Step-Audio-R1 from 71.1\% to 74.8\%, achieving the best overall result in the table. On MMSU, SKAD improves Qwen2.5-Omni-7B from 57.7\% to 61.4\%, and improves Step-Audio-R1 from 69.3\% to 71.6\%. These improvements indicate that SKAD can benefit both general audio understanding and spoken-language reasoning involving linguistic, semantic, prosodic, and paralinguistic cues.

\begin{table}[htp]
\setlength{\tabcolsep}{3pt}
\setlength{\extrarowheight}{0.5pt}
\centering
\small
\caption{Accuracies (\%) on MMAU (v05.15.25) and MMSU benchmarks. \textsuperscript{\textdagger} represents results reproduced in this work. \textbf{Bold} and \underline{underlined} indicate the best and the second best results among all models, respectively.}
    \begin{tabular}{@{}lcccc@{}|ccc}
        \hline
        \toprule
        \multirow{2}{*}{\textbf{Models}} & \multicolumn{4}{c}{\textbf{MMAU Test-mini}} & \multicolumn{3}{c}{\textbf{MMSU}} \\
        \cmidrule(lr){2-5} \cmidrule(lr){6-8}
        & Sound & Music & Speech & Avg. & Percept. & Reason. & Avg.\\
        \hline
        \multicolumn{5}{l}{\textit{\textbf{Baselines:}}} &  \\
        Gemini 1.5 Pro & 71.2 & 65.3 & \textbf{75.1} & 70.5 & 46.1 & 76.2 & 60.7 \\
        Gemini 2.0 Flash & 71.2 & 65.3 & \textbf{75.1} & 70.5 & - & - & - \\
        GPT-4o Audio & 64.6 & 56.3 & 66.7 & 62.5 & 39.7 & 72.0 & 56.4 \\
        Kimi-Audio & 75.7 & 66.8 & 62.2 & 68.2 & 43.5 & 76.0 & 59.3\\
        Audio Flamingo 3 & \underline{79.6} & \underline{74.0} & 66.4 & 73.3 & - & - & 61.4 \\
        \hline
        \multicolumn{5}{l}{\textit{\textbf{Ours:}}} \\
        Qwen2.5-Omni-7B\textsuperscript{\textdagger} & 78.7 & 64.4 & 67.9 & 70.3 & 44.3 & 72.1 & 57.7 \\
        \quad + SKAD & \textbf{83.2} & 67.1 & 70.9 & \underline{73.7} & 48.6 & 75.0 & 61.4 \\
        Step-Audio-R1\textsuperscript{\textdagger} & 73.6 & 68.6 & 71.2 & 71.1 & \underline{60.5} & \underline{78.6} & \underline{69.3} \\
        \quad + SKAD & 77.8 & \textbf{74.3} & \underline{72.4} & \textbf{74.8} & \textbf{61.3} & \textbf{82.6} & \textbf{71.6} \\
        \hline
    \end{tabular}
    \label{MMAUMMSU}
\end{table}

Overall, the main benchmark results show that SKAD provides stable improvements across different evaluation settings and model types. The gains on Qwen2.5-Omni-7B suggest that the proposed method can provide useful inference-time guidance for a general omni-modal model. The gains on Step-Audio-R1 further indicate that even a reasoning-oriented LALM with stronger built-in deliberation can still benefit from the proposed sketch-guided audio-contrastive decoding strategy.

\subsection{Ablation Study}

\begin{table*}[t]
\centering
\renewcommand{\arraystretch}{1.05}
\setlength{\tabcolsep}{3.5pt}
\footnotesize

\caption{
Ablation study on Qwen2.5-Omni-7B and Step-Audio-R1 across MMAR, MMAU, and MMSU.
Direct CoT denotes the original reasoning baseline, while Direct CD applies audio-contrastive decoding directly without constructing the acoustic evidence sketch.
SKAD denotes the proposed sketch-guided audio-contrastive decoding method.
Avg. is the average accuracy over the three benchmarks.
}
\label{tab:ablation}

\resizebox{\linewidth}{!}{%
\begin{tabular}{lc*{8}{c}}
\toprule
\multirow{2}{*}{\textbf{Variant}}
& \multirow{2}{*}{$\boldsymbol{\alpha}$}
& \multicolumn{4}{c}{\textbf{Qwen2.5-Omni-7B}}
& \multicolumn{4}{c}{\textbf{Step-Audio-R1}} \\
\cmidrule(lr){3-6}
\cmidrule(lr){7-10}
&
& \textbf{MMAR}
& \textbf{MMAU}
& \textbf{MMSU}
& \textbf{Avg.}
& \textbf{MMAR}
& \textbf{MMAU}
& \textbf{MMSU}
& \textbf{Avg.} \\
\midrule

Direct CoT
& --
& 53.4
& 70.3
& 57.7
& 60.5
& 62.6
& 71.1
& 69.3
& 67.7 \\

\midrule

\multirow{2}{*}{Direct CD}
& 0.2
& --
& --
& --
& --
& 62.2
& 71.3
& 69.4
& 67.6 \\

& 0.5
& 53.9
& 70.6
& 57.9
& 60.8
& --
& --
& --
& -- \\

\midrule

\multirow{4}{*}{SKAD}
& 0.1
& 54.5
& 73.4
& 60.9
& 62.9
& 64.1
& 74.5
& 71.5
& 70.0 \\

& 0.2
& 55.1
& 73.5
& 61.3
& 63.3
& \textbf{64.8}
& \textbf{74.8}
& \textbf{71.6}
& \textbf{70.4} \\

& 0.5
& \textbf{56.5}
& \textbf{73.7}
& \textbf{61.4}
& \textbf{63.9}
& \textbf{64.8}
& 74.7
& 71.0
& 70.2 \\

& 1.0
& 56.3
& \textbf{73.7}
& 61.0
& 63.7
& 64.2
& 74.5
& 69.9
& 69.5 \\

\bottomrule
\end{tabular}%
}
\end{table*}

Table~\ref{tab:ablation} presents the ablation results on Qwen2.5-Omni-7B and Step-Audio-R1 across MMAR, MMAU, and MMSU. We study two key factors: whether the acoustic evidence sketch is used before contrastive decoding, and how the contrastive strength $\alpha$ affects performance. Direct CoT denotes the original reasoning baseline. Direct CD applies contrastive decoding directly to the reasoning process without constructing an acoustic evidence sketch, while SKAD performs contrastive decoding after the sketch is generated.

First, directly applying contrastive decoding without the acoustic sketch does not bring consistent or substantial improvements. On Qwen2.5-Omni-7B, Direct CD with $\alpha=0.5$ only marginally improves the average accuracy from 60.5\% to 60.8\%. On Step-Audio-R1, Direct CD with $\alpha=0.2$ yields a comparable result, slightly changing the average accuracy from 67.7\% to 67.6\%. These results indicate that Direct CD provides limited and unstable gains, with its performance remaining largely comparable to the Direct CoT baseline. This observation supports our motivation that a directly audio-masked reasoning branch does not necessarily serve as a reliable contrastive reference. Without a shared acoustic sketch, the masked branch lacks sufficient audio-related context and may produce weak or unstable token distributions for contrastive decoding.

In contrast, SKAD consistently improves over Direct CoT under all tested values of $\alpha$. For Qwen2.5-Omni-7B, SKAD improves the average accuracy from 60.5\% to 62.9--63.9\%. For Step-Audio-R1, it improves the average accuracy from 67.7\% to 69.5--70.4\%. These results indicate that the acoustic evidence sketch plays an important role in making the contrastive branch informative. By providing both branches with a shared textual representation of the audio, SKAD makes the sketch-only branch a more meaningful reference and allows the audio-conditioned branch to be contrasted more effectively.

The optimal contrastive strength differs across backbone models. Qwen2.5-Omni-7B obtains the best average performance with $\alpha=0.5$, while Step-Audio-R1 performs best with $\alpha=0.2$. This difference is consistent with the characteristics of the two models. Qwen2.5-Omni-7B is a general omni-modal model and benefits from stronger contrastive guidance. Step-Audio-R1 is specifically designed for long-form audio reasoning, so a milder intervention is sufficient and avoids over-perturbing its original reasoning behavior. When $\alpha$ is increased to 1.0, the performance of Step-Audio-R1 drops from the best average accuracy of 70.4\% to 69.5\%, suggesting that overly strong contrastive intervention can weaken the model's built-in reasoning capability.

Overall, the ablation study shows that the effectiveness of SKAD depends on both the acoustic evidence sketch and the subsequent audio-contrastive decoding. Direct contrastive decoding without the sketch provides limited or negative gains, whereas sketch-guided contrastive decoding yields stable improvements across benchmarks and models. These findings support the design choice of applying contrastive decoding after constructing a shared acoustic evidence sketch rather than directly intervening in plain CoT reasoning.

\begin{figure}[t]
\centering
\includegraphics[width=12cm]{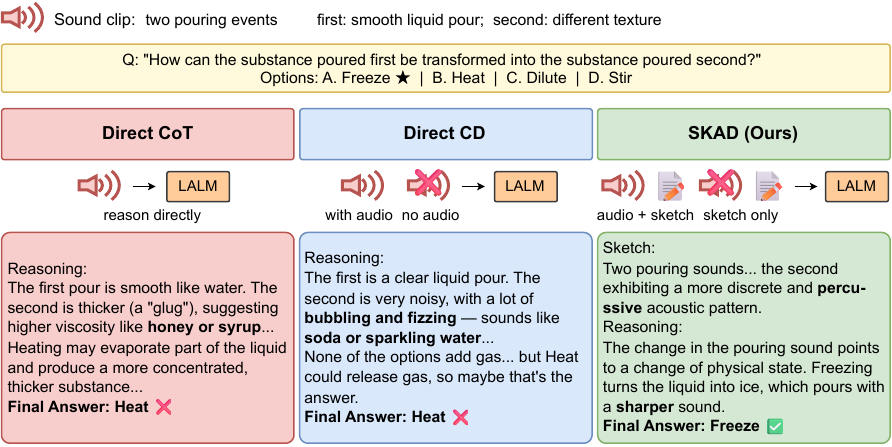}
\caption{Qualitative comparison of Direct CoT, Direct CD, and SKAD on an audio reasoning example.
Direct CoT drifts from the observed acoustic change to an unsupported viscosity-based hypothesis.
Direct CD contrasts audio-conditioned reasoning with a severely underconstrained audio-masked branch, making the contrastive reference unreliable.
SKAD first builds an acoustic evidence sketch as a shared semantic scaffold, and then contrasts the audio-conditioned branch with the sketch-only branch, encouraging predictions that are more specifically supported by the original audio.}
\label{case}
\end{figure}

\subsection{Case Analysis}

Fig.~\ref{case} presents a qualitative comparison of the reasoning behaviors of the three methods. Direct CoT incorrectly interprets the second pouring sound as evidence of a thicker liquid and consequently selects \textit{Heat}. Direct CD also predicts \textit{Heat}, misattributing the acoustic change to bubbling or fizzing. In contrast, SKAD first captures the more discrete and percussive characteristics of the second sound in its acoustic evidence sketch, which supports the inference of a change in physical state and leads to the correct answer, \textit{Freeze}. Overall, this example qualitatively suggests that the shared acoustic evidence sketch can provide a better-aligned and more informative reference for contrastive decoding, helping the model make more effective use of task-relevant acoustic cues during reasoning.

\section{Conclusion}

In this paper, we studied how to improve inference-time audio reasoning in Large Audio-Language Models through sketch-guided contrastive decoding. We observed that directly eliciting long-form reasoning can make the generation process increasingly affected by language-continuation bias, while applying contrastive decoding directly without intermediate audio evidence provides an unreliable masked reference branch. To address this issue, we proposed SKAD, a training-free Sketch-Guided Audio-Contrastive Decoding method. SKAD first constructs an acoustic evidence sketch to provide a shared semantic context, and then contrasts an audio-conditioned reasoning branch with a sketch-only branch during decoding. This design makes the contrastive reference more informative and helps the model emphasize predictions that are more strongly supported by the input audio.

Experiments on MMAR, MMAU, and MMSU show that SKAD consistently improves the average accuracy of both Qwen2.5-Omni-7B and Step-Audio-R1. The gains on Qwen2.5-Omni-7B indicate that the proposed method can provide useful audio-grounding guidance for a general omni-modal model, while the gains on Step-Audio-R1 show that even a reasoning-oriented LALM can benefit from a calibrated inference-time decoding strategy. Ablation studies further confirm that directly applying contrastive decoding without the acoustic evidence sketch is less effective, and that the contrastive strength should be adapted to the reasoning characteristics of different backbone models. Overall, our results suggest that effective audio reasoning requires not only eliciting intermediate reasoning, but also constructing a reliable contrastive reference that helps keep the decoding process grounded in acoustic evidence.

\subsubsection{Generative AI Use Disclosure.}
Generative AI tools were used only for English language polishing and \LaTeX{} formatting; they were not used to produce scientific claims, experimental results, or substantive content.

\begingroup\small\subsubsection*{\small Disclosure of Interests.} The authors have no competing interests to declare that are relevant to the content of this article.\par\endgroup

\bibliographystyle{splncs04}
\bibliography{mybib}

\end{document}